\documentclass[article]{JHEP4}
\usepackage{graphicx}
\usepackage{multirow}
\usepackage{cite}
\newcommand{\superiso}{Mahmoudi:2007vz,Mahmoudi:2008tp}
\newcommand{\higgsbounds}{Bechtle:2008jh,Bechtle:2011sb}

\usepackage{hep}
\preprint{CP3-15-05, MCnet-15-04}

\title{Higgs and Z boson associated production via gluon fusion in the SM and the 2HDM}

\author{B.~Hespel, F.~Maltoni, E.~Vryonidou\\
Centre for Cosmology, Particle Physics and Phenomenology (CP3),\\
 Universit\'e catholique de Louvain, B-1348 Louvain-la-Neuve, Belgium \\}
\abstract
{We analyse the associated production of Higgs and $Z$ boson via heavy-quark loops at the LHC in the Standard Model and beyond. We first review the main features of the Born $2\to 2$ production, and in particular discuss the high-energy behaviour, angular distributions and $Z$ boson polarisation. We then consider the effects of extra QCD radiation as described by the $2 \to 3$ loop matrix elements, and find that they dominate at high Higgs transverse momentum.  We show how merged samples of 0-- and 1--jet multiplicities, matched to a parton shower can provide a reliable description of  differential distributions in $ZH$ production.  In addition to the Standard Model study, results in a  generic two-Higgs-doublet-model are obtained and presented for a set of representative and experimentally viable benchmarks for $Zh^0$, $ZH^0$ and $ZA^0$ production. We observe that various interesting features appear either due to the resonant enhancement of the cross-section or to interference patterns between resonant and non-resonant contributions.}

\maketitle

\begin{document}

\section{Introduction}
\label{sec:intro}

Run I of the LHC has brought the discovery~\cite{Aad:2012tfa,Chatrchyan:2012ufa}
of a scalar particle, so far consistent with the Higgs boson predicted by the Standard Model
(SM)~\cite{Weinberg:1975gm}, and it has given the first evidence of the  Brout-Englert-Higgs
mechanism~\cite{Englert:1964et,Higgs:1964pj} in particle physics.  The increased energy and luminosity that will be achieved in Run II at the LHC will allow us to pin down the properties and in particular the strength and form of the interactions of such a boson with all other SM particles. To this aim a vast campaign of measurements of rates and distributions in various production and decay channels is being planned.

Among such processes is the associated production of a Higgs boson together with a vector boson $V$, either a $W^\pm$ or a $Z$, also known as Higgs-strahlung, {\it i.e.}, at the leading order in QCD, the Drell-Yan production of an off-shell vector boson $q \bar q \to V^*$ with its subsequent decay $V^* \to VH$. While suppressed in the SM with respect to the leading gluon-gluon and vector boson fusion channels, $VH$ production is of phenomenological interest mostly because the presence of the vector boson (and possibly of leptons coming from its decay) in the final state can help to access the large yet challenging $H\to b \bar b$ decay mode.  For instance, Higgs-strahlung has been the dominant Higgs search mode at the Tevatron\cite{TEVNPH:2012ab}. At the LHC, the ATLAS \cite{Aad:2014xzb} and the CMS \cite{Chatrchyan:2013zna} collaborations have investigated  $VH$ production, with the Higgs boson decaying to a $b-$quark pair, both reporting small excesses above the background only hypothesis. Searches for Higgs decaying to $W^+W^-$  \cite{ATLAS-CONF-2013-075,CMS-PAS-HIG-13-017} and to invisible states \cite{ATLAS-CONF-2013-011,CMS-PAS-HIG-13-018} have also been performed by both ATLAS and CMS. 

On the theory side, predictions for $ZH$ production are known at NNLO in QCD and at NLO electroweak in EW theory. 
The NNLO QCD cross section includes the Drell-Yan type terms of $\mathcal{O} (g^4 \alpha_s^2)$ first computed in  \cite{Hamberg:1990np,Brein:2003wg}. In addition to Higgs-strahlung, it has been noted that contributions from quark--anti-quark initiated diagrams where the Higgs is emitted from a top quark loop arise at the same order. These diagrams interfere with the LO and NLO Drell-Yan amplitudes and have been computed  in \cite{Brein:2011vx}, where they were found to contribute to the inclusive NNLO cross section at the percent level. Implementations of the NNLO QCD calculations are publicly available in {\sc vh@nnlo} \cite{Brein:2012ne} and {\sc HVNNLO} \cite{Ferrera:2014lca}. Fully differential NLO QCD and EW results  can be obtained with the program HAWK \cite{Ciccolini:2003jy,Denner:2014cla}, while event generation accurate at NLO in QCD (inclusively and for higher jet-multiplicitites), can be nowadays obtained (automatically or semiautomatically) in several frameworks, {\it i.e.},  {\sc MadGraph\_aMC@NLO} \cite{Alwall:2014hca} / {\sc POWHEG} \cite{Alioli:2010xd} + {\sc {\sc Pythia 8} \cite{Sjostrand:2007gs} /Herwig++}\cite{Bahr:2008pv} and {\sc Sherpa} \cite{Gleisberg:2008ta}. 

At NNLO, a purely virtual gluon fusion contribution emerges, through the $gg\rightarrow ZH$ amplitude squared, which at the LHC can be enhanced by the large gluon-gluon luminosity at small Bjorken $x$. Its contribution to the total cross section has been known for a long time~\cite{Dicus:1988yh,Kniehl:1990iva}  and it has been included in the implementations of the NNLO calculations~\cite{Ferrera:2014lca,Brein:2012ne}. The gluon fusion component is separately gauge invariant, IR and UV finite and accounts for about 10\% of the total NNLO cross section at 14~TeV.  Being essentially a leading-order contribution,  $gg\rightarrow ZH$ introduces a rather strong scale dependence to the NNLO result, which in turn is known quite precisely. In order to reduce the associated theoretical uncertainty, recently, NLO corrections for the gluon fusion contribution have been estimated by computing them in the infinite top-quark mass limit~\cite{Altenkamp:2012sx}. The NLO corrections to this process, $\mathcal{O}(\alpha_s^3)$, while formally part of the N$^3$LO $ZH$ cross section, are expected to be large, similarly to other gluon fusion processes such as Higgs single or pair production. The computation of the approximate NLO corrections in the infinite top mass limit has confirmed this expectation. The NLO computation in the infinite top mass limit reduces the scale uncertainty by a factor of two, yet the size of the finite top-quark mass effects remains unknown: the exact NLO result requires two-loop multi-scale amplitudes whose analytic form is beyond the current advances in the multi-loop technology.  In an effort to further reduce the theoretical uncertainties in this process, a soft gluon resummation for the gluon-gluon contribution has been performed in \cite{Harlander:2014wda} promoting the previous results to NLO+NLL accuracy. We should note here that in contrast with single Higgs production, where the infinite top-quark mass limit provides a good description of the process, and allows the computation of higher order corrections, here, similarly to (yet with even less control than) Higgs pair production, the much higher scales involved make the effective field-theory approach unreliable, especially so at the differential level. 

In addition to the SM production mechanism and characteristics, interesting features can be expected from Higgs production in association with a $Z$ boson in beyond the SM scenarios. The Two-Higgs-Doublet-Model (2HDM) is an attractive framework in which Higgs-strahlung can lead to interesting features.  First the range of channels is richer: in addition to the production of the light (125~GeV) Higgs boson in association with a $Z$ boson ($Zh^0$), $Z$ associated production of the heavy scalar ($ZH^0$) and pseudoscalar boson ($ZA^0$) are also possible \cite{Harlander:2013mla}. Experimental searches are already underway to look for signals of these processes, especially in the case where the cross-sections can be enhanced by the resonant production of an intermediate scalar ($H^0$ or $A^0$) with subsequent decays into $Z$ and a lighter scalar. In particular, CMS has searched for signals of the decay  of the pseudoscalar $A^0$ into a $Zh^0$ pair \cite{CMS-PAS-HIG-14-011} and that of the heavy scalar $H^0$ into a $ZA^0$ pair \cite{CMS-PAS-HIG-13-025} and the results have been used to set constraints on the 2HDM parameter space. 

So far considerable effort has been devoted to provide accurate total rates in this channel  for both the SM and the 2HDM, but accuracy and precision in the differential distributions is also of vital importance. This need becomes more important for experimental analyses which make use of exclusive observables, in order to tame the typically very large QCD backgrounds. Moving in that direction, it has been noted in the literature \cite{Englert:2013vua,Harlander:2013mla} that the gluon induced component can play an important role. The gluon fusion Higgs $p_T$ distribution peaks at higher values than the corresponding Drell-Yan one, and therefore its relative contribution becomes more important in boosted Higgs searches, which are preferred experimentally to reduce the backgrounds. The prospects of such searches have improved recently due to progress in jet substructure techniques, after the seminal suggestion in \cite{Butterworth:2008iy}.

The aim of this work is to contribute to the understanding of gluon induced $ZH$ production and to improve the predictions for the differential distributions. We consider the $2\to 2$ and $2\to3$ matrix elements entering the gluon fusion contribution to $Z$ Higgs-strahlung. We first review the main features of the $2\to 2$ ones and then examine the importance of the $2\to3$ contributions. Given the lack of an exact and fully differential NLO computation for this process, we  provide a better description of the kinematics for this component by combining the $2\to 2$ and $2\to3$ matrix elements  in a merged sample, matched to a Parton Shower (PS). This provides a fully exclusive control at the hadron level. A similar approach has been followed for other loop induced processes, such as single Higgs production \cite{Alwall:2011cy} and Higgs pair production \cite{Li:2013flc,Maierhofer:2013sha}. In general, this method provides a better description of the kinematics, yet as the formal accuracy for total rates remains at LO, it is often combined with a normalisation obtained from higher-order computations, when available.

This merging-matching approach makes use of the fact that while tree level fixed-order amplitudes describe reliably the region of hard and well separated jets, the parton shower  provides a better description of the soft and collinear regions. Combining the two requires of course a consistent treatment to avoid double-counting, which is achieved by various merging algorithms.  Methods that are widely used for tree level merging are CKKW \cite{Catani:2001cc,Krauss:2002up}, CKKW-L \cite{Lonnblad:2001iq,Lavesson:2005xu} (and their later improvements \cite{Hoeche:2009rj,Hamilton:2009ne}), and MLM \cite{Mangano:2006rw}. More recently new methods have been developed to perform the merging at NLO, see for example FxFx \cite{Frederix:2012ps} and UNLOPS \cite{Lonnblad:2012ix}, yet not directly applicable to $2\to 2$ loop-induced processes at the Born level yet, mainly due the absence of analytic results for the two-loop $2 \to 2$ matrix elements.  

In this work we study gluon induced Higgs-strahlung at the LHC,  presenting the first merged-matched results for $g g \to Z\phi$, with $\phi$ being a generic scalar,  by employing the 0 and 1-jet matrix elements for the SM and the 2HDM. This paper is organised as follows. In Section 2, we discuss the process within the SM, first by reviewing the important features coming from the $g g \to Z H $  matrix elements. We also consider the behaviour of the $2\to 3$ matrix elements, which we then combine with the $2 \to 2$ ones. We describe our methodology, and present results both at the parton level and after merging and matching to a parton shower. In Section 3, we explore the results of various 2HDM scenarios using the same calculation setup. We draw our conclusions in the final section. 

\section{Gluon induced $ZH$ production in the SM}

\begin{figure}[h!]
\centering
\includegraphics[scale=0.5]{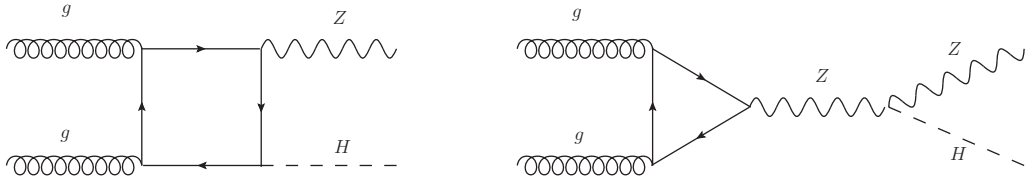}
\caption{Representative Feynman diagrams for $ZH$ production in gluon fusion in the SM.}
\label{diagrams0}
\end{figure}

Representative Feynman diagrams contributing to the $g g \to Z H$ process in the SM are shown in Fig.~\ref{diagrams0}. Massive fermions, $t$ and $b-$quarks, run in the box, while all flavours run in the triangle. The contribution of the two light generations to the triangle vanishes as required by the anomaly cancellation. In practice,  it is only the axial vector part of the heavy-quark-$Z$ coupling that contributes to the amplitude. The amplitude for this process was first computed in \cite{Dicus:1988yh,Kniehl:1990iva}. 

\begin{figure}[h!]
\centering
\includegraphics[scale=0.4]{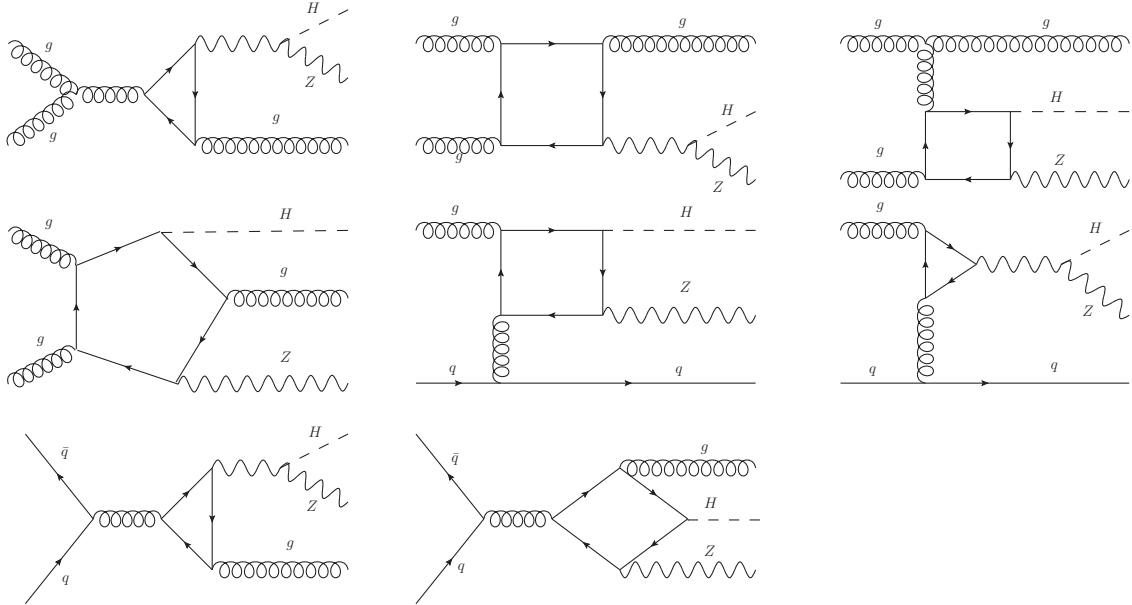}
\caption{Representative Feynman diagrams for gluon induced $ZHj$ production in the SM.}
\label{diagrams1}
\end{figure}

In what follows, we will first review the main features of the $2\to 2$ process for gluon induced $ZH$ production before discussing the implications of the $2\to 3$ one. A sample of the relevant diagrams contributing to $ZHj$ is shown in Fig.~\ref{diagrams1}. In addition to the $gg$ initial state amplitudes, the $qg$ and $q\bar{q}$ channels also open up, when an additional jet is allowed. The $gg \to  Z H  g$ amplitudes were used in \cite{Agrawal:2014tqa} to calculate the $gg$ part of the $ZHj$ cross-section at the LHC for various jet transverse momentum cuts. In what follows, we will consider these along with the $qg$ and $q\bar{q}$ diagrams to discuss the behaviour of the $2 \to 3$ amplitudes and subsequently to obtain a merged sample of 0 and 1-jet multiplicitities. 

\subsection{Calculation setup} 
In this work, we employ the {\sc MadGraph5\_aMC@NLO} framework \cite{Alwall:2014hca}. The one-loop amplitudes squared for $ZH$ and $ZHj$ can be obtained with the help of {\sc MadLoop} \cite{Hirschi:2011pa}, which computes one--loop matrix elements using the {\sc OPP} integrand--reduction method~\cite{Ossola:2006us} (as implemented in {\sc CutTools}~\cite{Ossola:2007ax}). A reweighting procedure is then employed to overcome the present limitations concerning event generation for loop-induced processes~\footnote{Automated event generation for loop-induced processes is currently being finalised\cite{loop-induced}.}. A reweighting method has been employed already for a series of processes within the {\sc MadGraph5\_aMC@NLO} framework \cite{Alwall:2011cy,Hespel:2014sla,Maltoni:2014eza} both at LO and NLO accuracy. This procedure involves generating events through the implementation of a tree-level effective field theory (EFT), in this case obtained by taking the limit of infinite top-quark mass with all other quarks being massless.  In practice, a UFO model \cite{Degrande:2011ua,deAquino:2011ub} including the effective theory interactions is imported in the simulation framework. After event generation, event weights obtained from the tree-level EFT amplitudes are modified by the ratio of the full one-loop amplitude over the EFT ones, {\it i.e.}, $r=|\mathcal{M}_{Loop}^2|/|\mathcal{M}_{EFT}^2|$, where $|\mathcal{M}_{Loop}^2|$ represents the numerical amplitude as obtained from {\sc MadLoop}. In our case, reweighting proves to be efficient in terms of the  computational speed, as the loop amplitudes have to be calculated for significantly fewer phase-space points than what would be needed to integrate them directly. Moreover the EFT leads to distributions that are in general harder in the tails, and therefore the EFT events populate regions that are later suppressed by the exact loop matrix elements, resulting to no significant degradation of the statistical uncertainty. 

\subsection{Parton level results}

Before proceeding to the technical setup and presenting results of the merging-matching,  we consider the salient aspects as observed at the parton level. The findings of this study will reveal some previously unnoticed features of $gg \to ZH$ and will act as a motivation to employ a merging-matching procedure in the following section. 

\begin{figure}[h]
\centering
\includegraphics[scale=0.6]{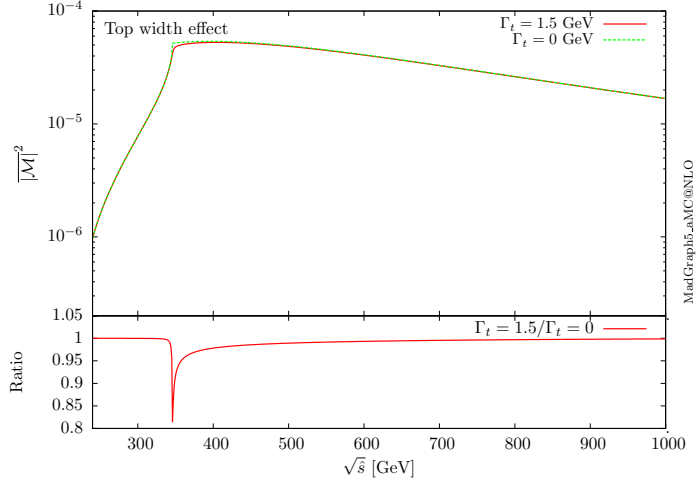}
\caption{Top width effect on the matrix element squared for $gg\to ZH$. Results for $\Gamma_t$=0 and 1.5~GeV are shown along with their ratio.}
\label{msquared}
\end{figure}

In our computation the heavy quark masses are set to: $m_t=$173 GeV and $m_b=$4.75 GeV, while the Higgs mass to $m_H=$125 GeV and the heavy quark Yukawas are given by $y_q/\sqrt{2}=m_q/v$. We note here that finite width effects in the propagators of the loops can be taken consistently into account within {\sc MadGraph5\_aMC@NLO} via the implementation of the complex mass scheme \cite{Denner:1999gp,Denner:2005fg}.  The effect of a non-zero top width is shown in Fig.~\ref{msquared}, where the matrix element squared for $gg\to ZH$, for 90$^0$ scattering, is shown as a function of the invariant mass of the $ZH$ system.  The correction is more important at the $t\bar{t}$ threshold, where it reaches 20\%. Finally, when integrated over all centre-of-mass energies and scattering angles, we find the top-quark width to modify the $g g \to ZH $ cross-section by $\sim$2\% at 14TeV, an effect similar to that observed for single and double Higgs production in \cite{Anastasiou:2011pi} and \cite{Maltoni:2014eza}, respectively. For the rest of the results presented in this work the width of the top  quark is set to zero. 

\begin{figure}[h!]
\centering
\includegraphics[trim=0.5 0 0 0, scale=0.75]{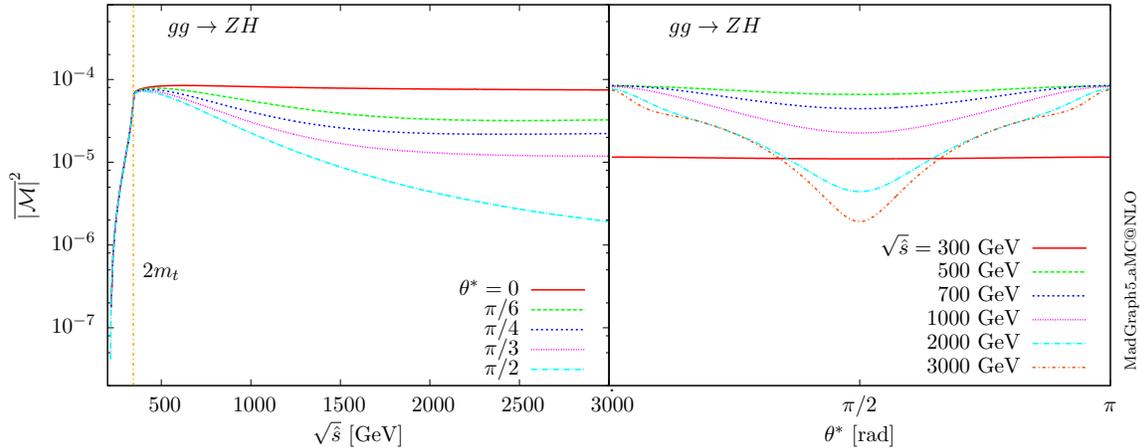}
\caption{Matrix element squared for $gg\to ZH$ as a function of the centre-of-mass energy for different values of scattering angles  (left) and as a function of the centre-of-mass scattering angle for various values of $\sqrt{\hat{s}}$ (right). }
\label{msquared_angle}
\end{figure}

An interesting aspect of the $g g \to ZH $ matrix element is its angular dependence. While in Fig. \ref{msquared} we have fixed the scattering angle to 90$^0$, in Fig. \ref{msquared_angle}, we show the dependence of the amplitude squared on the centre-of-mass scattering angle, for various values of $\hat{s}$. The matrix element starts with no angular dependence at low energies, but varies significantly with the angle at high energies. This angular dependence of the matrix element implies that at high energies, very forward or backward scattering is favoured over 90$^0$ scattering. This behaviour originates from the interplay between the triangle and the box diagrams, and their respective angular behaviour. As we will also discuss later, box and triangle interfere destructively, with the triangle contribution dominating at low energies. The cancellation becomes nearly exact at high energies, mostly leaving a remainder from the box contribution that is strongly dependent on the scattering angle.

We now proceed to discuss results for the LHC. For these, parton distribution functions (PDFs) are evaluated using the MSTW2008LO set \cite{Martin:2009iq} and the central renormalisation and factorisation scales are set to the invariant mass of the $ZH$ system: $\mu^0=\mu^0_R= \mu^0_F=m_{ZH}$.   {\sc MadGraph5\_aMC@NLO} allows the automatic computation of the scale and PDF uncertainties by a reweighting procedure \cite{Frederix:2011ss}. In our results, scale variations are obtained by varying the scales in the range of $\mu^0/2< \mu_{R,F} < 2\mu^0 $.

Table \ref{tab:sigma} summarises  results for the $gg \to ZH$ cross section and the ``loop-induced" $ZHj$ contribution, originating from the square of the amplitudes shown in Fig.~\ref{diagrams1}. For reference we also include the total $p p \to ZH$ cross section at NNLO obtained with {\sc vh@nnlo} \cite{Brein:2012ne}. The NNLO cross section includes the $gg \to ZH$ result of the first row, for which excellent agreement has been found between our computation and the result of {\sc vh@nnlo}. We note that the results shown in the second row of Table \ref{tab:sigma} for $ZHj$ are obtained using the loop amplitudes shown in Fig.~\ref{diagrams1}. These $qg$ and $q\bar{q}$ amplitudes can interfere with the Drell-Yan type real emission amplitudes. This interference contribution to the cross section has been computed in \cite{Brein:2011vx} and found to be at the per-mille level. In our computation we use these amplitudes squared, {\it i.e.}, at  $\mathcal{O}(\alpha_s^3)$. It is clear that at this order, other $qg$ loop-induced contributions can enter squared, for example, the set of diagrams where the $Z$ couples to a light quark and the $H$ to a top-quark loop. We have not included these diagrams here, as we consider them of a different origin, but we have checked that their amplitude squared contribution to the cross section is small, below the femtobarn level, and therefore at least one order of magnitude smaller than those in Fig.~\ref{diagrams1}. The interference of this type of diagrams with the Drell-Yan amplitude was   computed in \cite{Brein:2011vx} and  also found to be small. 

Given that the $ZHj$ amplitude is divergent in the limit of a collinear or soft jet,  we apply a cut  on the $p_T$ of the jet to obtain finite results. We have set this cut to 30~GeV in Table \ref{tab:sigma}. The $2 \to 3$ contribution comes mainly from the $gg$ initiated diagrams, with $qg$ giving about 20\% of the $ZHj$ cross section. The $ZHj$ contribution is not as suppressed as expected from the extra power of $\alpha_s$, leading to results comparable in size to the $gg\to ZH$ cross section. Of course these results are extremely sensitive to the chosen cut for the transverse momentum of the additional parton, as the cross section diverges in the IR limit. Such a problem would not arise in the case of a NLO computation matched to a parton shower, for example with the MC@NLO method \cite{Frixione:2002ik}, in which all divergences are regularised and cancelled for inclusive observables. 

\begin{table*}[t]
\renewcommand{\arraystretch}{1.3}
\begin{center}
    \begin{tabular}{l|c|c|c}
        \hline \hline
     Contribution [fb]  & $\sqrt{s} = 8$~TeV
         & $\sqrt{s} = 13$~TeV
         & $\sqrt{s} = 14$~TeV\\
         \hline
          $gg \to ZH$ &  
 17.4 $^{+34\%}_{-24\%}$ & 
58.5 $^{+30\%}_{-21\%}$ & 70.7 $^{+29\%}_{-21\%}$ \\         
        $pp \to ZHj$ ($p_T^j>30$ GeV)  & 12.4 $^{+52\%}_{-32\%}$ & 49.0 $^{+44\%}_{-32\%}$
& 58.4 $^{+47\%}_{-31\%}$ \\
$pp \to ZH $ (NNLO)& 387 $^{+2.2\%}_{-1.6\%}$ & 795 $^{+3.2\%}_{-2.0\%}$ & 886 $^{+3.2\%}_{-2.3\%}$ \\
\hline \hline
\end{tabular}
 \caption{\label{tab:sigma} Cross sections (in fb) for $ZH$ associated production
   at the LHC at $\sqrt{s} =  8, 13$ and 14~TeV. 
      The  uncertainties (in percent) refer to scale variations.  
      No cuts are applied to final state particles apart from the jet $p_T$ cut in the second row ($p_T^j >30$ GeV) and no Higgs or $Z$ branching ratios are included. The $ZHj$ contribution shown here is obtained from the loop diagrams shown in Fig.~2, while the NNLO results are obtained with {\sc vh@nnlo} \cite{Brein:2012ne}.}  
\end{center} 
\end{table*}

The results in Table \ref{tab:sigma} also demonstrate the problem of large scale uncertainties for the LO $gg$ cross section, that contribute significantly to the total NNLO scale uncertainty. The problem persists also for the loop-induced $ZHj$ contribution. A significant reduction of these intrinsic QCD uncertainties can only be achieved by a complete NLO computation, as discussed in \cite{Altenkamp:2012sx}.    

\begin{figure}[h!]
\centering
\includegraphics[scale=0.8]{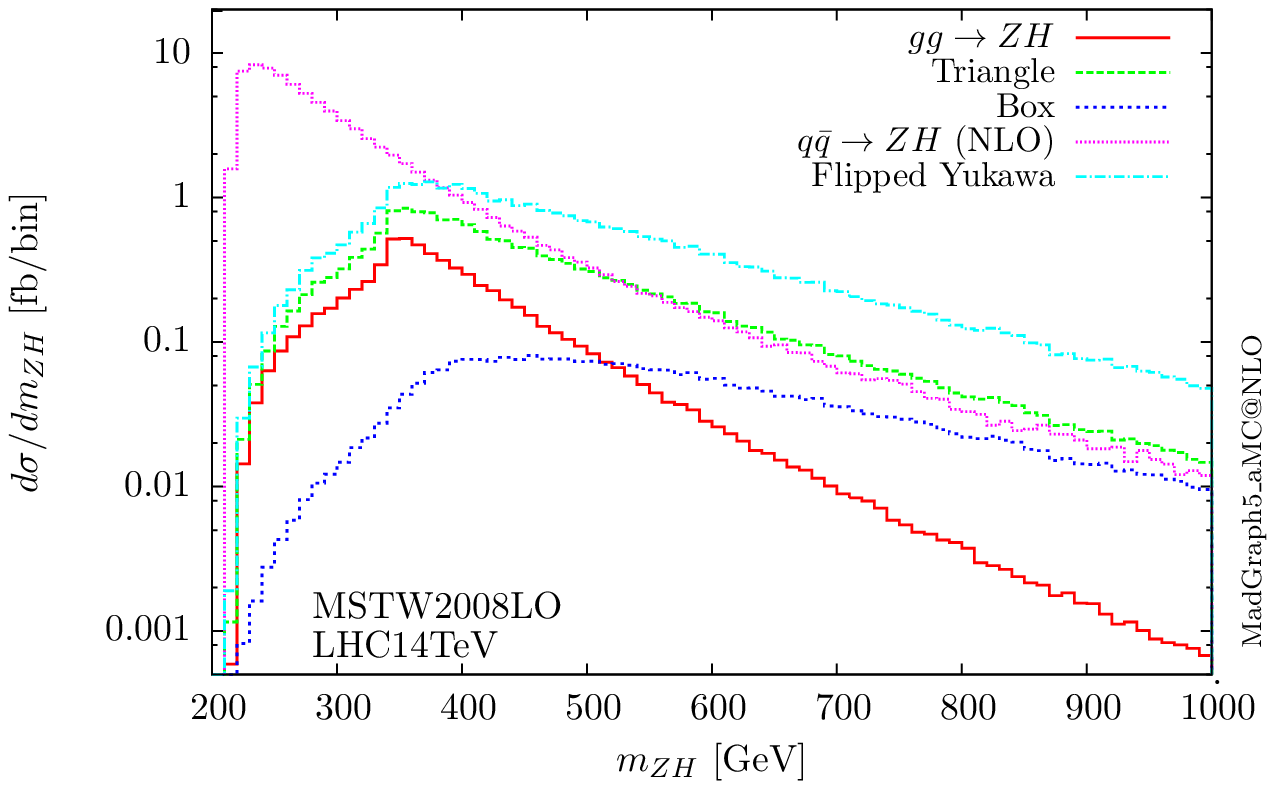}
\includegraphics[scale=0.8]{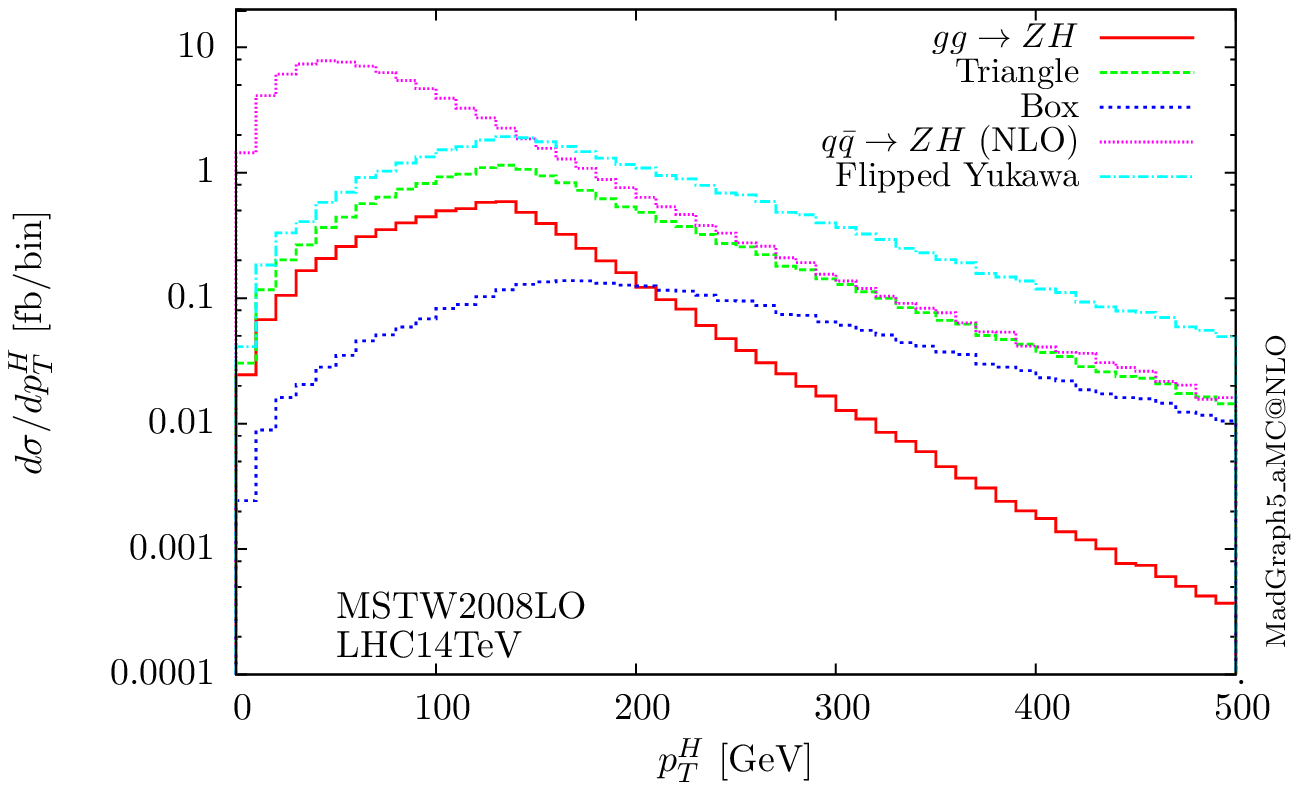}
\caption{Invariant $ZH$ mass and $p^H_T$ distributions for $ZH$ production at $\sqrt{s}=14$ TeV.  The gluon fusion contribution is decomposed into the triangle and box contribution. For completeness we also plot the Drell-Yan type contributions at NLO obtained with {\sc MadGraph5\_aMC@NLO}.
}
\label{mzhpth}
\end{figure}

In addition to the total cross-section results presented in Table \ref{tab:sigma}, interesting observations can be made by studying the differential distributions for the gluon fusion process. We start by presenting  distributions for the invariant mass of the $ZH$ system and the transverse momentum of the Higgs in Fig.~\ref{mzhpth} for the LHC at 14 TeV. These distributions have been shown elsewhere in the literature, for example in \cite{Englert:2013vua}, yet we consider them here again for completeness. In addition to the gluon fusion results, we also include the NLO Drell-Yan like distributions obtained automatically with  {\sc MadGraph5\_aMC@NLO}. We note here that differential NNLO results for the Drell-Yan like contribution can be provided by the code {\sc HVNNLO}. As is evident from \cite{Ferrera:2014lca}, the NNLO computation leads to an overall 20\% decrease of the Drell-Yan component but not to any significant shape difference compared to the corresponding NLO one. 

The first observation regards the clear presence of the $2m_t$ threshold, at which the gluon fusion amplitude acquires an absorptive part, related to the on-shell $gg \to t\bar t$ , $t \bar t \to ZH$ scattering, leading to a characteristic rise in the invariant mass distribution. It is evident from Fig.~\ref{mzhpth} that the gluon fusion component leads to distributions of fundamentally different shape from the Drell-Yan ones and therefore it should be considered in all relevant studies, in particular in the boosted region of $p_T^H>100$ GeV, where its relative importance increases.

In the plots, we decompose the gluon fusion result into the triangle and box components. The two interfere destructively over the whole range of centre-of-mass energies, with the cancellation between the two being nearly exact at high energies. Such a cancellation is due to unitarity: while each of the two diagrams grows with energy the cancellation leads to a well-behaved amplitude at high energies. We stress here that this behaviour of the amplitude is not present in the infinite top mass limit. In this limit, the amplitude for the box diagram vanishes and therefore only the triangle contributes to the amplitude, giving a rather bad approximation of the one-loop amplitude at high energies. In addition, we note that this is a process highly sensitive to the relative phase between the $HZZ$ and $Ht\bar{t}$ couplings. To demonstrate this, in Fig.~\ref{mzhpth} we show the result obtained by changing the relative sign between the top Yukawa and the $HZZ$ coupling. In pair with other processes where such unitarity cancellations take place, such as $H \to \gamma \gamma$ or $pp \to tHj$ \cite{Maltoni:2001hu,Biswas:2012bd,Farina:2012xp}, flipping the sign results in an increase in the gluon fusion induced contribution by a factor of five, and much harder distributions as the interference between triangle and box becomes constructive, see Fig.~\ref{mzhpth}. We conclude that, given the difference in the shape as well as the size of the cross section  above 2$m_t$, the $ZH$ invariant mass or transverse momentum of the  Higgs or the $Z$ could also be used to bound the relative phase between the Higgs couplings to fermions and to vector bosons.

\begin{figure}[h!]
\centering
\includegraphics[scale=0.8]{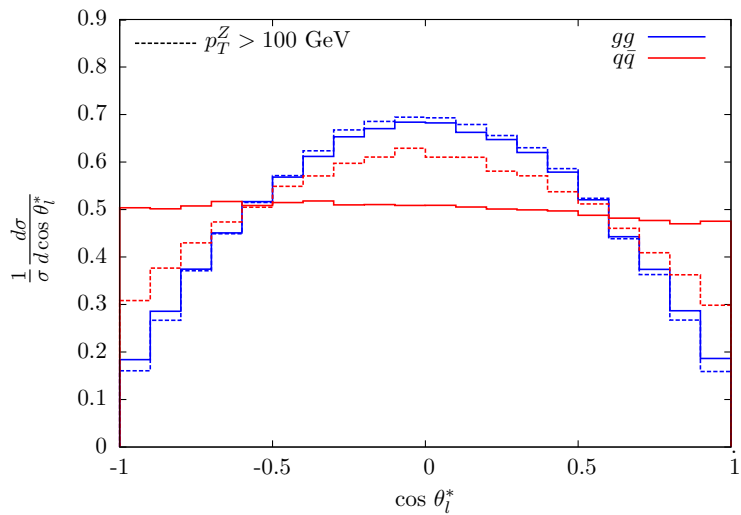}
\caption{Normalised distributions of the lepton angle $\theta^*_l$ for gluon fusion and Drell-Yan like $Z H$ production at $\sqrt{s}=14$ TeV. Results for an imposed cut of 100~GeV on the $p_T$ of the $Z$ are shown in dashed lines.}
\label{anglelepton}
\end{figure}

The difference in the $p_T$ shape between the Drell-Yan and gluon fusion production persists also in the distribution of the lepton $p_T$ coming from the $Z$ decay. Besides, another interesting aspect of the gluon fusion process is that it leads to different angular distributions for the resulting leptons compared to the Drell-Yan component \cite{Maltoni:2013sma}.  This is evident from studying the normalised distributions of the angle $\theta^*_l$, defined as the angle of the lepton between the lepton and $Z$ direction in the $Z$ rest frame is shown in Fig. \ref{anglelepton}. In the plot, we use the NLO Drell-Yan result, and plot the distributions with and without a cut of 100~GeV on the $p_T$ of the $Z$. The shape of the distribution without any cut, is significantly different, with the tree-level $ZH$ giving a flat distribution while the gluon-fusion one peaks at 90$^0$. The shape becomes similar for $p_T^Z>100$ GeV, while a 200 GeV cut (not shown here) completely eliminates the difference. This behaviour is related to the polarisation of the $Z$ which differs between the two production modes. This can be quantified by examining the relevant polarisation fractions in Table~\ref{tab:polarisationfraction}, as these are defined in \cite{Stirling:2012zt}. The fact that the $Z$ in $gg\to ZH$ is predominantly longitudinal leads to the central peak, while the small difference between $f_L$ and $f_R$ leads to a very mild asymmetry for $q\bar{q}$. Setting a 100~GeV cut on the $Z$ $p_T$ changes these values in agreement with the equivalence theorem, {\it i.e.}, by increasing the longitudinal polarisation fraction. For completeness, we also mention here that the main background for this process, $Z+b$-jets leads to predominantly left-handed $Z$ bosons \cite{Bern:2011ie,Stirling:2012zt} and therefore to different angular distributions, that could be potentially used as an additional discriminating handle to distinguish signal and background.


\begin{table}[tb!]
\begin{center}
 \begin{tabular}{l|c|c|c} \hline \hline
 Process & $f_0$ ($\%$) & $f_L$ ($\%$) & $f_R$ ($\%$) \\ \hline 
$g g \rightarrow Z H$ & 82.2 & 8.9 & 8.9  \\
$g g \rightarrow Z H$, $p_T^Z>100$ GeV & 86.3 & 6.9 & 6.8  \\ \hline
$q \bar{q} \rightarrow Z H$ & 35.6 & 32.4 & 32.0   \\
$q \bar{q} \rightarrow Z H$, $p_T^Z>100$ GeV & 62.6 & 18.8 & 18.6   \\ \hline \hline
 \end{tabular}
 \end{center}
 \caption{Polarisation fractions for the gluon fusion and NLO Drell-Yan production mode of $ZH$ at 14~TeV with and without a cut on the $Z$ $p_T$.}
 \label{tab:polarisationfraction}
\end{table}

\begin{figure}[h!]
\centering
\includegraphics[scale=0.8]{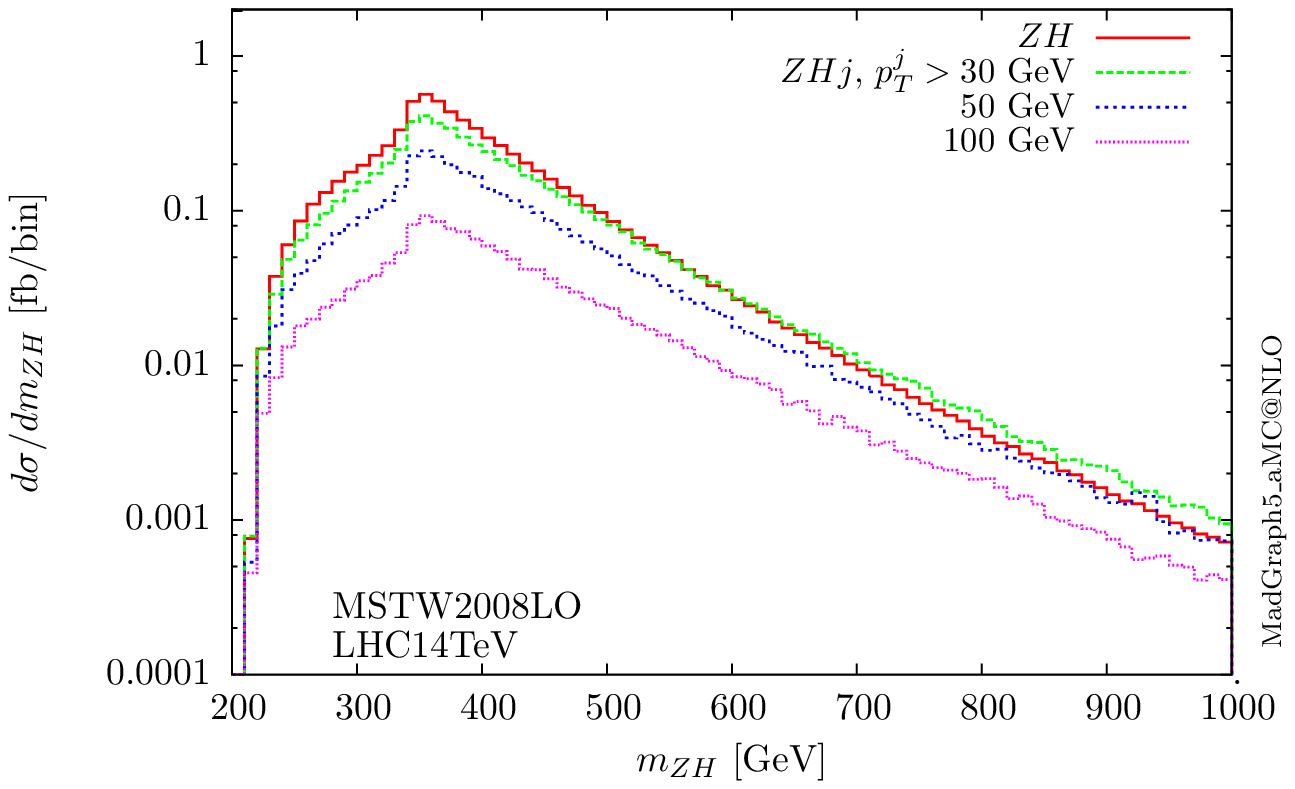}
\includegraphics[scale=0.8]{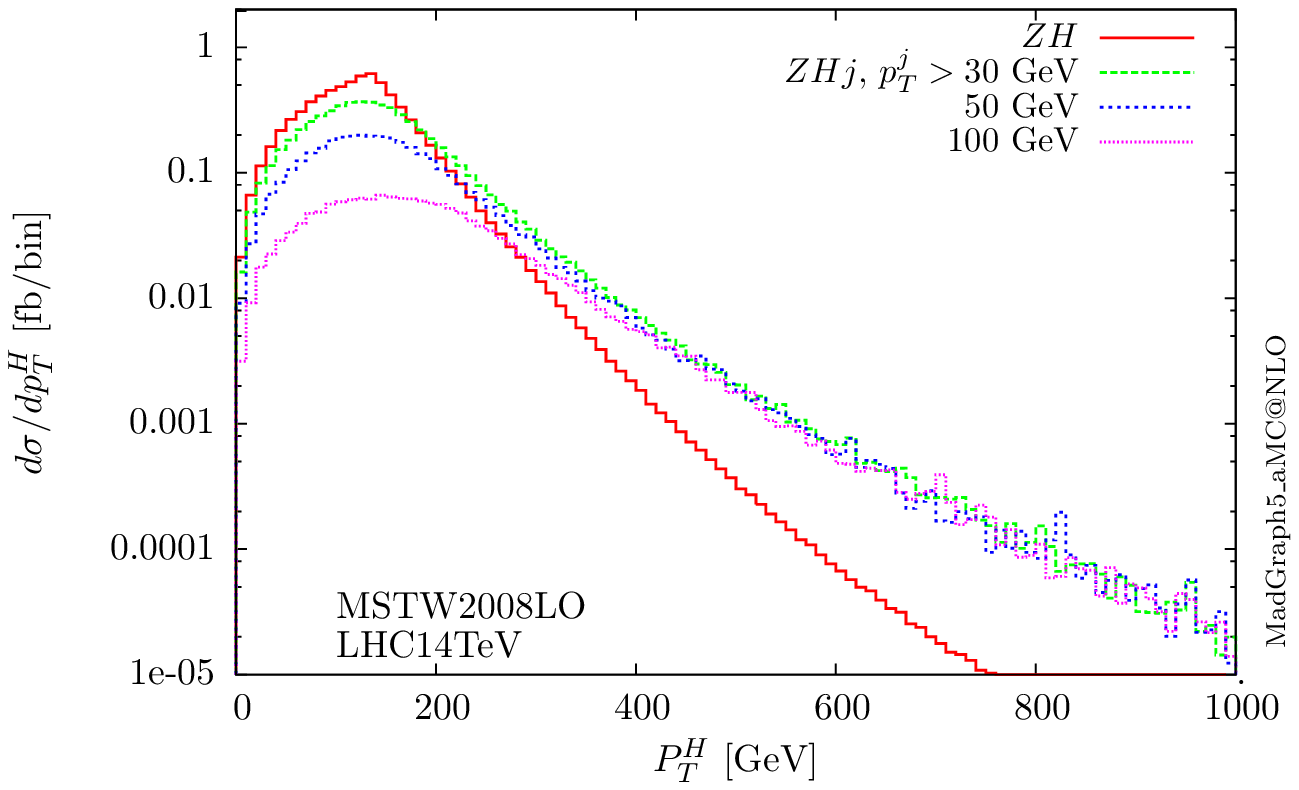}
\caption{Invariant $ZH$ mass and $p^H_T$ distributions for loop-induced $Z H$ and $ZHj$ production at $\sqrt{s}=14$ TeV. The results for the $ZHj$ distributions are shown for various jet $p_T$ cuts: 30, 50 and 100~GeV and again concern only the loop diagrams of Fig.~2. }
\label{mzhpthjet}
\end{figure}

Further to the $gg\to ZH$ results that we have discussed above, interesting conclusions can be drawn by studying the loop-induced $ZHj$ distributions. We have seen in Table \ref{tab:sigma} that these contributions are not negligible and their relative importance  increases with the centre of mass energy. A complete NLO computation for $gg \to ZH$ would be fully inclusive in these contributions but as such a computation is not available yet, we aim to draw some conclusions by studying them independently. For such a study a minimum cut has to be set on the transverse momentum of the additional jet to avoid the divergent  soft and collinear limit. We compare the distributions of the invariant mass of $ZH$ and the $p^H_T$ to those from  $2\to 2$ amplitudes by varying the $p_T$ cut set on the additional jet in Fig.~\ref{mzhpthjet}. 

The $ZH$ invariant mass distribution  shows that for all values of the jet $p_T$ cut the bulk of the cross-section remains close to the $2m_t$ threshold. The characteristic threshold behaviour due to the absorptive part of the amplitude remains visible for all cuts. At high invariant masses, we find that the amplitudes with an extra parton fall more slowly and overtake the $2 \to 2$ process. 

In contrast to the invariant mass distribution where no extreme modification of the shape takes place, the $p_T^H$ distribution is very much affected. First, we note that the threshold corresponding to $m_{ZH}\sim 2m_t$ is now not visible in the $p_T$ distributions for $ZHj$. The second and more striking observation is that above 300 GeV the $2 \to 3$ process leads to a much harder  $p_T$ spectrum compared to the  $2\to2$ one. Moreover, for $p^H_T$ values above 400 GeV all three distributions for $ZHj$ coincide. The fact that in this region the result is insensitive to the jet $p_T$ cut implies that hard jet emissions are dominating. This occurs because by allowing the emission of a parton new kinematic configurations open up. In this high $p_T$ region, the kinematic configuration in which a soft jet is emitted and the $Z$ and $H$ basically recoil against each other is not the most favourable one. Instead, the configuration in which a hard jet recoils against the $H$, with the $Z$ remaining rather soft becomes the preferred one. We have explicitly confirmed this behaviour by setting a high cut on the $p_T$ of the Higgs, and studying the corresponding jet and $Z$ transverse momentum distributions. A clear preference for the configurations where the jet is hard and the $Z$ is rather soft is found when sufficiently far from the IR divergence. The behaviour of the $2 \to 3$ amplitudes at high $p_T$ can be traced back to the presence of $t-$channel gluon diagram such as the $gg\to ZHg$ one shown in the top right of Fig.~\ref{diagrams1}, which becomes dominant in this region. The same behaviour is displayed by the $qg\to ZH q$ contributions, when these are considered separately, as they include diagrams of the same type as shown in the second row of Fig.~\ref{diagrams1}. 

In conclusion, we have found that, especially for the transverse momentum distribution of the Higgs, the emission of an additional jet can dramatically modify the shape, due to new allowed kinematic configurations. This effect might prove important in studies involving highly boosted Higgs as discussed for example in \cite{Englert:2013vua}. The $2 \to 3$ matrix elements are important and therefore need to be taken into account for accurate simulations. To combine the two in a consistent way and therefore provide a more realistic picture of the differential distributions, we will resort to merging and matching to a parton shower. In the following section we will discuss how this method allows us to provide more accurate predictions for the distributions.  

\subsection{Merging different jet multiplicities: setup}

Given the lack of a complete NLO computation and the relevance of the $2 \to 3$ matrix elements, the best available procedure to accurately predict the distribution shapes is to employ the Matrix-Element--Parton Shower (ME+PS) procedure. ME+PS  schemes allow the consistent combination of matrix elements with different jet multiplicities via their matching to a parton shower. In our study we employ the $k_T$-MLM scheme as implemented in {\sc MadGraph5\_aMC@NLO}. Merged samples are then passed to {\sc Pythia 8}  \cite{Sjostrand:2007gs,Sjostrand:2014zea} for PS. 

The implementation of MLM merging in {\sc MadGraph5\_aMC@NLO}/{\sc Pythia 8} comes in two variants: the traditional $k_T$-MLM and the shower-$k_T$ schemes. The two give comparable results as discussed in \cite{Alwall:2008qv}. In this study we will employ the shower-$k_T$ scheme. While this scheme has been used for phenomenological studies with {\sc Pythia 6} in the past, see for example \cite{deAquino:2012ru}, we employ  the most recent implementation of the scheme in {\sc MadGraph5\_aMC@NLO} combined with {\sc Pythia 8}. For a detailed description of this approach one can refer to \cite{Alwall:2008qv}, while here we only mention its main points. 

In the  shower-$k_T$ scheme, matrix element events are generated with a minimum separation $p_{T_{\textrm{min}}}$,  between parton and the initial state ($iB$), and $Q_{\textrm{cut}}$ between final-state partons ($ij$), defined by the measure:
\begin{equation}
d^2_{iB}= p^2_{T_i}> p^2_{T_{{\rm min}}},\,\,\,\,    d^2_{ij}=\textrm{min}(p_{T_i}^2,p_{T_j}^2)\Delta R^2_{ij} > Q^2_{\rm cut},
\end{equation} where $\Delta R^2_{ij}=2[\textrm{cosh}(\eta_i-\eta_j)-\textrm{cos}(\phi_i-\phi_j)]$ and $p_{T_i}$, $\eta_i$ and $\phi_i$ are the transverse momentum, pseudorapidity and azimuthal angle of particle $i$. Short distance (parton level) events are then passed to {\sc Pythia 8} which evolves them down using the $p_T$-ordered shower. In practice, for each event {\sc Pythia 8} records the scale of the hardest shower emission: $Q^{PS}_{\rm{hardest}}$. This scale is then used to accept or reject the event as follows: for the low multiplicity events, the event is rejected if $Q^{PS}_{\rm{hardest}}>Q_{\rm{cut}}$, while for the highest multiplicity the event is rejected if $Q^{PS}_{\rm{hardest}}>Q^{ME}_{\rm{softest}}$, with $Q^{ME}_{\rm{softest}}$ being the scale of the softest matrix-element parton in the event.  The value of $Q_{\rm{cut}}$ is selected on a process-by-process basis to ensure that there is a smooth transition between the ME and PS regimes. In practice, this is assessed by examining the differential jet rate distributions which show if the transition is indeed smooth.

\subsection{Merged sample results for $ZH$ in gluon fusion}

Using the setup described in the previous subsection for the merging and matching, we present in this section our merged results for various observables. In our simulations we keep the $H$ and $Z$ stable. For the merging performed here, the shower-$k_T$ scheme is used with $Q_{\textrm{cut}}=p_{T_{\textrm{min}}}=30$ GeV. We have checked that this choice leads to smooth differential jet rate distributions, and therefore a smooth transition between the ME and PS regimes. 

\begin{figure}[h!]
\centering
\includegraphics[trim=1.5cm 0 0 0,scale=0.88]{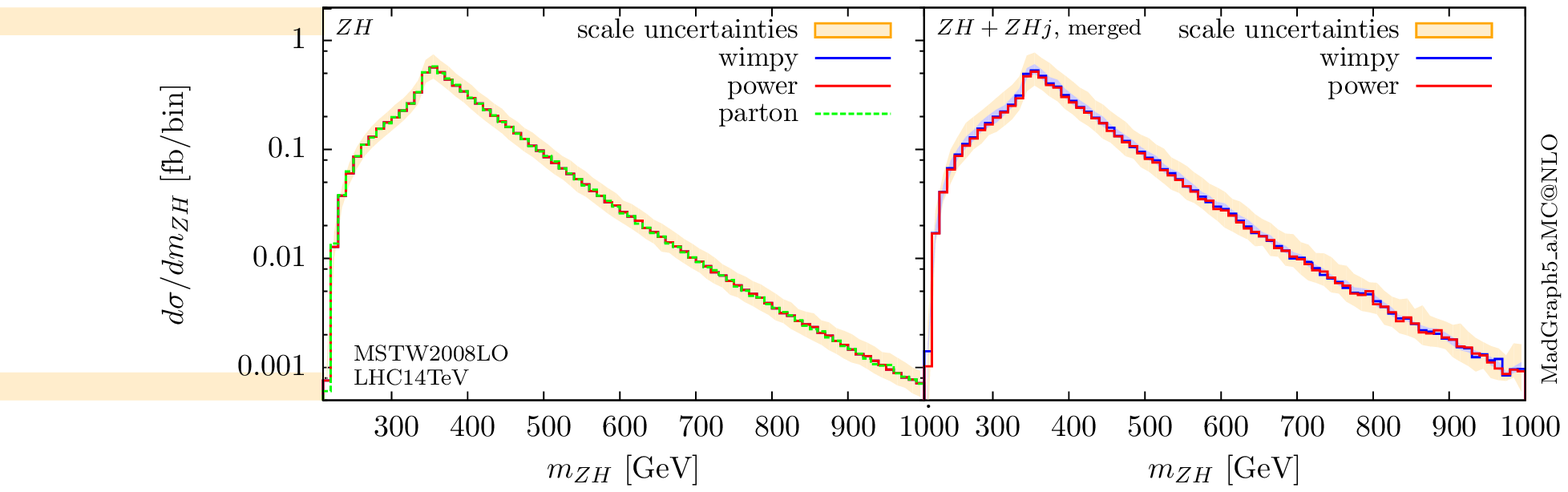}
\includegraphics[trim=1.5cm 0 0 0,scale=0.88]{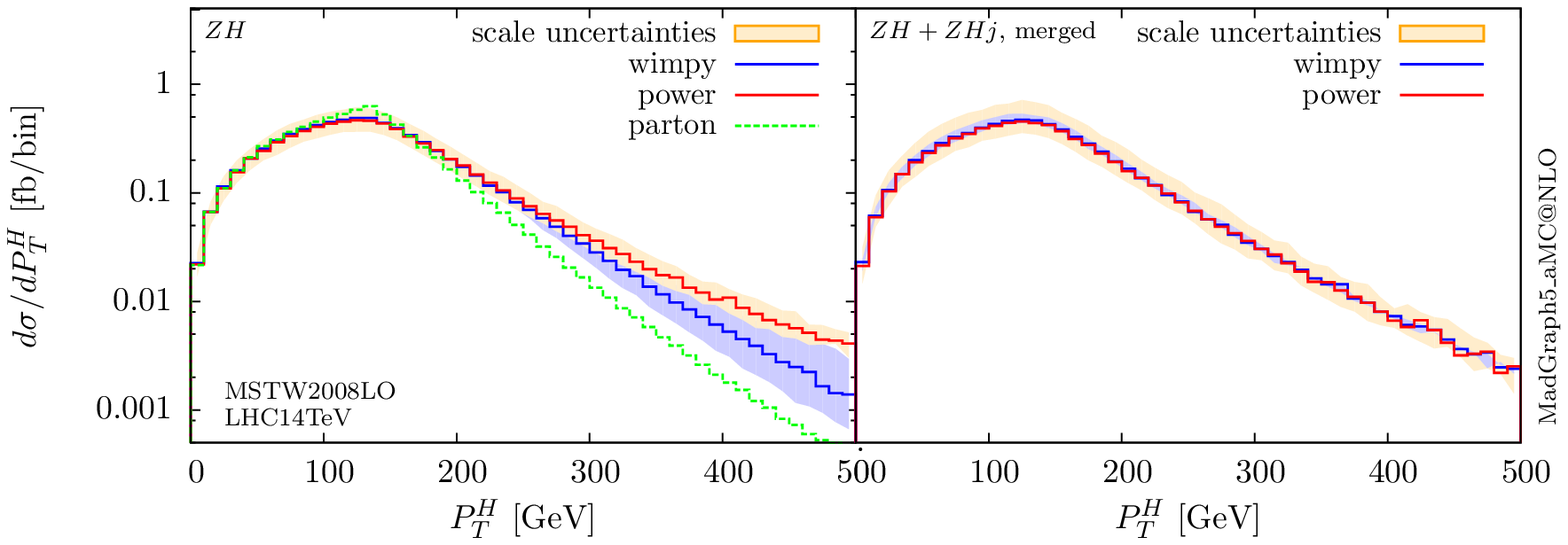}
\caption{Invariant $ZH$ mass and $p^H_T$ distributions for gluon induced $ZH$ production at $\sqrt{s}=14$ TeV. The left column shows the results obtained for the $gg\to ZH$ case, with different starting scale for the shower: ``wimpy" and ``power'' shower. The blue band shows the variation of the shower scale for ``wimpy" shower in the range $0.5 \mu_F< Q_{PS} <2\mu_F$, while the yellow bands show the uncertainty associated with a factor of two variation of the renormalisation and factorisation scales with respect to their central value. The right column shows the same results for the merged sample. The green curves in the left column correspond to the parton level results before passing them through {\sc Pythia 8}. }
\label{mzhpthmerged}
\end{figure}

\begin{figure}[h!]
\centering
\includegraphics[trim=1.5cm 0 0 0,scale=0.88]{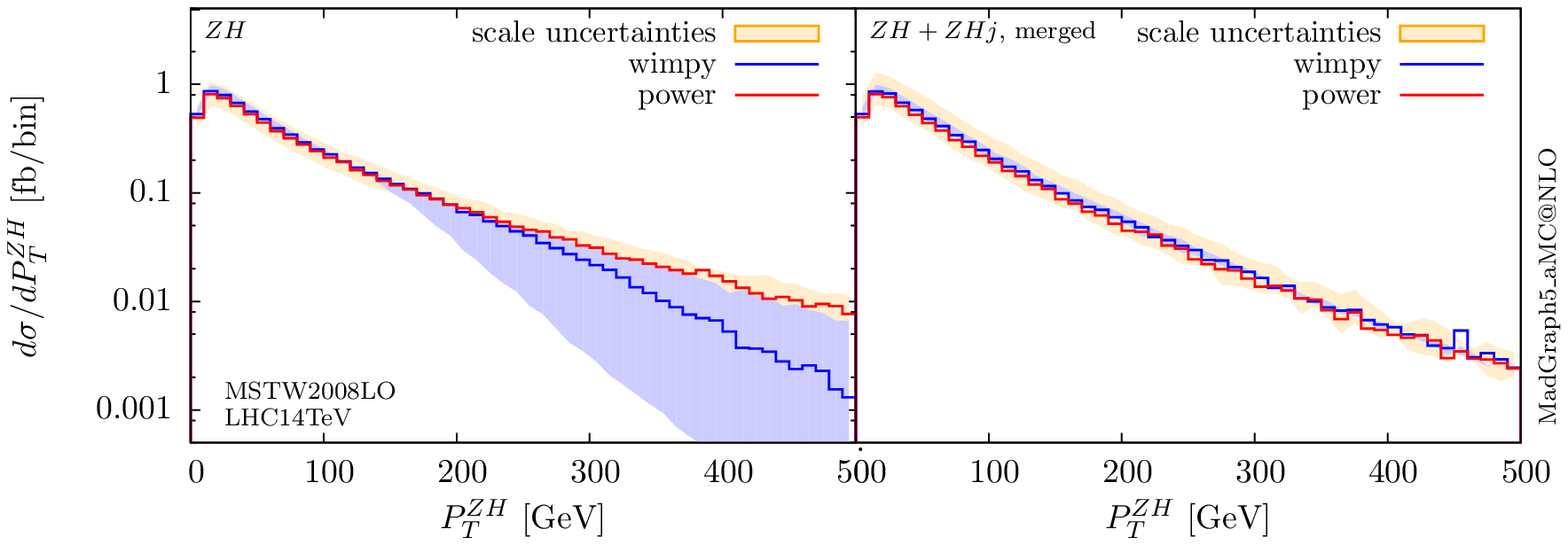}
\includegraphics[trim=1.5cm 0 0 0,scale=0.88]{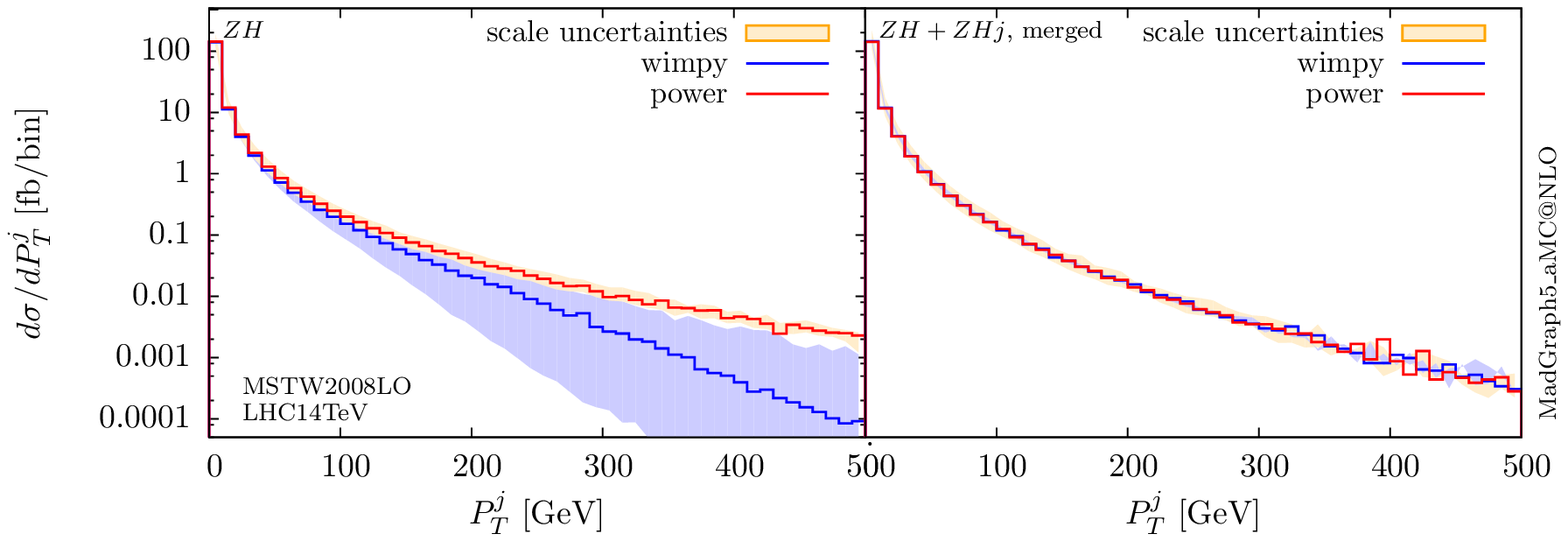}
\caption{Transverse momentum of the $ZH$ system and hardest jet $p_T$ distributions for $ZH$ production at $\sqrt{s}=14$ TeV. The setup is the same as in Fig.~8. }
\label{ptzhptjmerged}
\end{figure}

We start by presenting the results for the invariant mass of the $ZH$ system and the $p_T$ of the Higgs in Fig.~\ref{mzhpthmerged}, while  $p_T^{ZH}$ and $p_T^j$ distributions are shown in Fig.~\ref{ptzhptjmerged}.  A comparison is made between the $gg \to ZH$ sample showered with {\sc Pythia 8} and the merged 0 and 1-jet matched sample, presented in combination with the uncertainties associated with scale choices for both the factorisation/renormalisation scale of the hard process and the shower starting scale. We set the central value for the renormalisation and factorisation scales to $m_{ZH}$, as for the parton-level results. The shower starting scale in {\sc Pythia 8}  can be set to either the kinematical limit ($p_T= \frac{\sqrt{\hat{s}}}{2}  $), corresponding to what we refer to as ``power"--shower or the factorisation scale of each event ($m_{ZH}$ in our case), {\it i.e.}, ``wimpy"--shower. {\sc Pythia 8} allows us, for the ``wimpy"-shower case, to modify the shower starting scale in the range of $0.5 \mu_F< Q_{PS} <2\mu_F$. This gives us the possibility to  systematically study the dependence of the results on the choice of the shower scale for both the merged and $gg\to ZH$-only samples, as shown by the blue bands in the plots. To study the systematic uncertainties due to renormalisation and factorisation scale variations, we vary the scales between $0.5 \mu^0< \mu_{R,F} < 2 \mu^0$, with $\mu^0=m_{ZH}$. This variation is shown by the yellow bands in the plots (with the central prediction being the ``power"-shower result). In the results shown for $m_{ZH}$ and $p_T^H$ in Fig.~\ref{mzhpthmerged}, we also include the parton-level results for comparison purposes.

First, we notice that not all distributions are sensitive to the procedure of merging-matching. In particular, the invariant mass of the $ZH$ system shows no shape variation.
In this process, we only have initial state radiation and therefore significant changes in the shape are not expected for an observable like $m_{ZH}$. Other observables, on the other hand, are highly sensitive to the choice of shower parameters. The distributions for the transverse momentum of the Higgs, $p^H_T$, but more importantly the transverse momentum of the $ZH$ system, $p_T^{ZH}$, and that of the hardest jet,  $p_T^j$, which are trivially zero at parton-level, depend strongly on the shower parameters. We first notice that the shower produces a $p_T^H$ distribution harder than the parton-level one for all shower scale choices. This is related to the harder behaviour of the $2\to 3$ distributions discussed earlier. 

Another interesting observation to be made is related to the shape changes associated with the shower scale choice. The ``power"-shower leads to consistently harder distributions, while the ``wimpy"-shower gives softer distributions. The different shower predictions start to diverge in a region correlated with the invariant mass of the $ZH$ system, as this is the factorisation scale which is taken to be the starting scale of the ``wimpy"-shower. The shower scale uncertainty bands become wider at larger $p_T$ values. This is more evident in the second set of observables, $p_T^{ZH}$ and $p_T^j$, for which the non-merged predictions can vary by more than one order of magnitude between different shower scale options. At high transverse momentum, the shower uncertainty becomes more important than the intrinsic QCD one associated to the factorisation and renormalisation scale choice for the hard process. We note here that despite the fact that the factorisation and renormalisation scale uncertainty is large,  as evident from the yellow bands, it seems to mainly affect the normalisation of the curves.

The advantage of the ME+PS procedure is then made obvious by noticing that the shower scale uncertainty is almost completely eliminated in the merged predictions. For all observables, the shower scale uncertainty bands remain well within the corresponding renormalisation and factorisation scale uncertainty ones, even at high transverse momentum. ME+PS predictions are therefore more accurate/precise and predictive than the parton shower alone as they include the exact $2\to 3$ matrix elements. These play an important role in the phase space regions populated by highly boosted objects which is often the case for LHC searches. 

\section{Z$\phi$ production in the 2HDM}
In the previous section we discussed gluon induced $ZH$ production in the SM, employing the  ME+PS merging  method to improve the accuracy of the predictions for the differential distributions at the LHC.  In this section, we will follow a similar approach for a beyond the SM scenario. The case we consider in this work is the 2HDM, as a minimal extension of the SM \cite{Branco:2011iw}. The 2HDM extends the scalar sector of the SM by introducing a second $SU(2)_L$ doublet $\Phi_2$, which leads to five physical Higgs bosons, {\it i.e.} in the case of $CP$ conservation, the light $CP$-even one, $h^0$, a heavier  $CP$-even one, $H^0$, a  $CP$-odd one, $A^0$, and two charged Higgs bosons $H^{\pm}$. Assuming no extra sources of $CP$-violation, seven input parameters fully characterise the model:
\begin{equation}
\tan \beta, \sin \alpha, m_{h^0}, m_{H^0}, m_{A^0}, m_{H^{\pm}}, m^2_{12},
\end{equation}
with the convention $0\leq\beta-\alpha<\pi$ (with $0<\beta<\pi/2$) fixing the sign of the Higgs coupling to the gauge bosons to be the same as in the SM. Depending on the structure of the Yukawa couplings, two main types of 2HDM setups can be considered: type $I$ where all fermions couple to just one of the Higgs doublets and type $II$, where up-quarks (down) couple only to $\Phi_2$ ($\Phi_1$). 

Various theoretical requirements such as stability, perturbativity and unitarity, impose constraints on the 2HDM parameter space. At the same time, electroweak precision measurements and recent LHC Higgs physics results further constrain the parameter space, as discussed in more detail in \cite{Hespel:2014sla}. Nevertheless, 2HDM scenarios that satisfy these constraints and yet have a significantly different phenomenology than the SM exist and have been studied extensively in the literature. These scenarios arise in the ``decoupling'' limit \cite{Gunion:2002zf}, {\it i.e.}, in the limit of cos$(\beta-\alpha)\ll 1$, which ensures that the masses of the additional Higgs bosons lie well above the light-Higgs one. Even in this case, significant shifts from the SM couplings are allowed, in particular in the tan$\beta \gg 1$ limit, known as ``delayed decoupling'' \cite{Haber:2000kq}. Interestingly, the ``decoupling" limit is not the only 2HDM realisation which is consistent with all parameter space constraints. Scenarios with light additional Higgs bosons are also allowed in the so-called ``alignment limit" \cite{Gunion:2002zf}.

\begin{figure}[ht]
\includegraphics[scale=0.38]{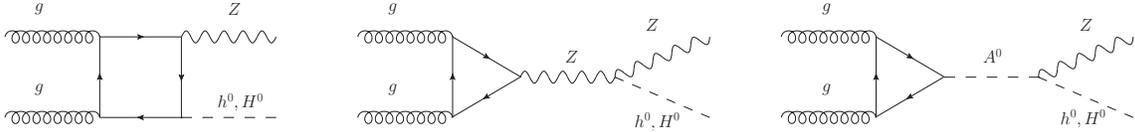}
\caption{Representative Feynman diagrams for $ZH^0$/$Zh^0$ production in gluon fusion in the 2HDM.}
\label{diagrams0b}
\end{figure}

\begin{figure}[h!]
\centering
\includegraphics[scale=0.4]{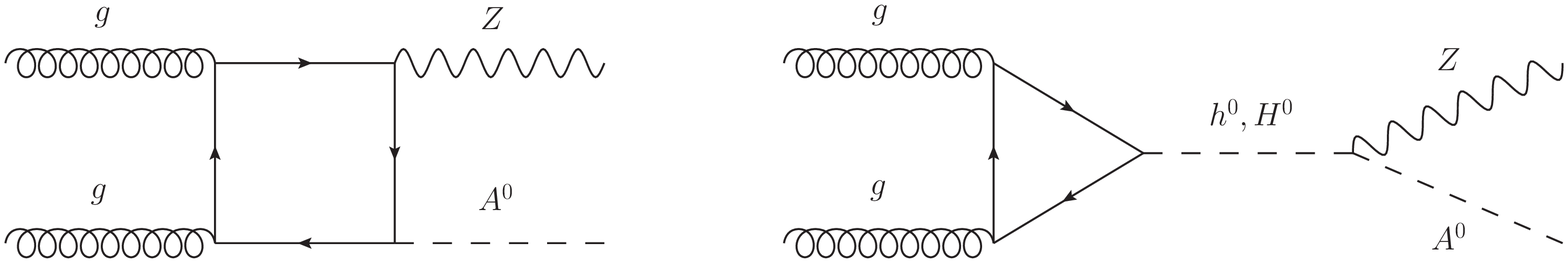}
\caption{Representative Feynman diagrams for $ZA^0$ production in gluon fusion in the 2HDM.}
\label{diagrams0c}
\end{figure}

Some examples of viable 2HDM scenarios will be considered in this section to explore possible 2HDM signatures in Higgs production in association with a $Z$ boson. Interesting features can arise in this process, not only because of possible deviations of the light Higgs couplings from their SM values, but most importantly because of the presence of the heavier states,  $H^0$ and  $A^0$, which can lead to resonant production of $Z \phi$ final states.  Three neutral combinations of final states are possible: $Zh^0$, $ZH^0$ and $ZA^0$. These 2HDM processes have already been discussed in \cite{Harlander:2013mla}. Similarly to $ZH$ production in the SM, the production of the $Zh^0$ and $ZH^0$ final states can occur through Drell-Yan type diagrams, and in gluon-gluon fusion. The Drell-Yan like cross sections can be obtained straightforwardly by the appropriate rescaling of the SM cross-sections by the ratio of the $g^{\phi}_{ZZ}$ coupling over its SM counterpart, but the situation for the gluon fusion case is more involved. This can be inferred by considering the corresponding Feynman  diagrams for the gluon fusion processes, shown in Figs.~\ref{diagrams0b} and \ref{diagrams0c}. The possibility of resonant production depends on the masses of $A^0$ and $H^0$, while interesting interference patterns can arise due to relative sign of the $A^0 \phi Z$ couplings. For completeness and to facilitate the discussion that follows, the dependance of the relevant Yukawa couplings on the 2HDM parameters is shown in Table \ref{tab:couplingsuY}, for type-$I$ and type-$II$ setups, as rescalings of their SM counterparts. We note that the following couplings are also relevant for these process  (valid for both type-$I$ and type-$II$ setups):
\begin{equation}
\hat{g}^{h^0}_{VV}=\sin(\beta-\alpha), \quad \hat{g}^{H^0}_{VV} = \cos(\beta-\alpha), \quad \hat{g}^{A^0}_{VV}=0, 
\label{HVV}
\end{equation} 
with $\hat{g}^{\phi}_{VV}$ being the rescaling of the $\phi VV$ coupling compared to the $HVV$ one in the SM, while the $A^0 \phi Z$ couplings are proportional to:
\begin{equation}
g^{A^0h^0}_Z = \cos(\beta-\alpha) \,\, \textrm{and} \,\, g^{A^0H^0}_Z = -\sin(\beta-\alpha).
\label{HZA}
\end{equation} 


\begin{table}[tb!]
\begin{center}
 \begin{tabular}{c|c|c} \hline \hline
 Coupling & type-$I$ & type-$II$  \\ \hline 
$\hat{g}_u^{h^0}$ & $\cos \alpha / \sin \beta$ & $\cos \alpha / \sin \beta$   \\
$\hat{g}_d^{h^0}$ & $\cos \alpha / \sin \beta$ & $-\sin \alpha / \cos \beta$  \\
$\hat{g}_u^{H^0}$ & $\sin \alpha / \sin \beta$ & $\sin \alpha / \sin \beta$  \\
$\hat{g}_d^{H^0}$ & $\sin \alpha / \sin \beta$ & $\cos \alpha / \cos \beta$  \\
$\hat{g}_u^{A^0}$ & $\cot \beta$ & $\cot \beta$  \\
$\hat{g}_d^{A^0}$ & $-\cot \beta$ & $\tan \beta$   \\ \hline \hline
 \end{tabular}
 \end{center}
 \caption{Dependence of Yukawa couplings for up and down-type quarks on the 2HDM parameters for type-$I$ and type-$II$ setups. The expressions in the table correspond to the ratio of the couplings over the corresponding SM value.}
 \label{tab:couplingsuY}
\end{table}

We stress here that several studies have been presented in the literature in particular for the $A^0Z$ process, mostly  in the context of the MSSM \cite{Kao:1991xg,Yin:2002sq,Kao:2003jw,Yang:2003kr,Kniehl:2011aa}.  In the case of the MSSM, there are more constraints on the values of the Higgs couplings, while the 2HDM allows more freedom that can lead to more striking signals. A particularly interesting cosmologically motivated 2HDM scenario leading to a $A^0\to ZH^0$ signature at the LHC is presented in \cite{Dorsch:2014qja}, that finds very good prospects for discovery or exclusion even for the low-luminosity LHC. We also mention that various 2HDM scenarios allow significantly enhanced bottom Yukawas, and the $Z\phi$ states can be produced mainly through $b\bar{b}$ annihilation. This has been extensively discussed in the literature \cite{Yang:2003kr,Kao:2004vp,Li:2005qna}, and in relation with the subtleties of the treatment of the bottom quarks \cite{Harlander:2011aa}. In this work, we will be focussing on the gluon fusion channel, presenting results for a series of 2HDM benchmarks.

\subsection{Calculation setup}
The calculation setup, regarding the reweighting and the ME+PS merging procedure, follows closely that described in the previous section for the SM. In this section we discuss the details specific to the 2HDM implementation. The computation is performed within the {\sc MadGraph5\_aMC@NLO} framework \cite{Alwall:2014hca}, where {\sc MadLoop}~\cite{Hirschi:2011pa} is used for the computation of the one--loop amplitudes. The 2HDM@NLO model obtained from the package {\sc NLOCT} \cite{Degrande:2014vpa} is imported into {\sc MadGraph5\_aMC@NLO} for the computation of the 2HDM amplitudes. The model is based on the {\sc FeynRules} \cite{Alloul:2013bka} and {\sc UFO }\cite{Degrande:2011ua} frameworks. More importantly for our computation, it includes all the necessary UV counterterms and R2 vertices for the {\sc MadLoop} calculation. The model allows the computation of tree--level and one--loop amplitudes within a completely general 2HDM setup. 
The 2HDM parameters for the different benchmarks  are imported into {\sc MadGraph5\_aMC@NLO} \cite{Alwall:2014hca} using a parameter card,  constructed with the help of an in--house modification of the public calculator {\sc 2HDMC} \cite{Eriksson:2009ws}, in the same way as in \cite{Hespel:2014sla}.

We stress again here that as described in \cite{Hespel:2014sla} the 2HDM benchmarks to be used here have been constructed in 
agreement with all up--to--date parameter space constraints, which we have included through an interface of the public tools {\sc 2HDMC}~\cite{Eriksson:2009ws},
{ \sc HiggsBounds}~\cite{\higgsbounds}, {\sc SuperIso}~\cite{\superiso} and {\sc{HiggsSignals}} \cite{Bechtle:2013xfa,Stal:2013hwa} along with 
additional routines of our own. 

Three benchmarks will be employed to present the 2HDM results, with the corresponding parameters shown in Table \ref{tab:benchmarks}. Benchmarks B1 and B2 have been constructed and used already for our study of Higgs pair production in the 2HDM \cite{Hespel:2014sla}, they correspond to B1 and B4 in \cite{Hespel:2014sla}. Benchmark B3 is a new one designed for this study. Here we briefly mention the main features of each benchmark.  For completeness we also show the couplings relevant for $gg\to Z\phi$ production in Table \ref{tab:couplingsYZ}, as rescalings of the SM couplings, similarly to Table \ref{tab:couplingsuY} and Eqs. \ref{HVV}-\ref{HZA}.

\begin{itemize}
 \item{\textbf{Benchmark B1}}: 
 A type-II 2HDM scenario with moderately heavy Higgs masses. Small tan$\beta$ and cos$(\beta-\alpha)$ values ensure that the couplings of the light Higgs boson remain SM-like. The bottom Yukawa is slightly enhanced. This scenario allows a resonant production of both the light and Heavy Higgs with a $Z$ boson through the decay of the pseudoscalar $A^0$. The sign of the $Zh^0A^0$ coupling determines the interference of the $A^0$-mediated production with the SM-like diagrams.
 
 \item{\textbf{Benchmark B2}}:
 A type-I 2HDM scenario with a relatively light heavy Higgs $H^0$ and a significantly heavier pseudoscalar $A^0$. Both light-Higgs top and bottom Yukawas are enhanced by $\sim10\%$.  The negative sign of $m_{12}^2$ protects the stability of the vacuum. This scenario also allows the resonant production of both the light and Heavy Higgs with a $Z$ boson through the decay of the pseudoscalar $A^0$.
 
\item{\textbf{Benchmark B3}}: 
 Another type-II 2HDM scenario with a reversed mass hierarchy between the heavy scalar $H^0$ and the pseudoscalar $A^0$. The small $\tan \beta$ value allows us not to over-suppress the $\hat{g}_{A^0 tt}$ coupling, while the $\hat{g}_{A^0 bb}$ is enhanced. Thanks to the inverted mass hierarchy $m_{h^0}<m_{A^0}<m_{H^0}$ the resonant production of  $A^0$ with a $Z$ boson due to the heavy neutral Higgs decay becomes kinematically allowed.
 
 \end{itemize}

\begin{table}[tb!]
\begin{center}
 \begin{tabular}{l|rr|rrrr} \hline \hline
 & $\tan\beta$ & $\alpha/\pi$ & $m_{H^0} $ &  $m_{A^0} $  & $m_{H^{\pm}} $   & $m^2_{12} $  \\ \hline 
B1 & 1.75 & -0.1872 & 300 & 441 & 442 & 38300  \\
B2 & 1.20 & -0.1760 & 200 & 500 & 500 & -60000 \\
B3 & 1.70 & -0.1757 & 350 & 250 & 350 & 12000  \\ \hline \hline
 \end{tabular}
 \end{center}
 \caption{Parameter choices for the different 2HDM benchmarks used in our study. All
 masses are given in GeV. The lightest Higgs mass is fixed in all cases to $m_{h^0} = 125$ GeV. }
 \label{tab:benchmarks}
\end{table}

\begin{table} [tb!]
\begin{center}
    \begin{tabular}{  l | r | r | r | r | r | r || r | r | r | r }
    \hline \hline
      & $\hat{g}_{h^0 tt}$ & $\hat{g}_{h^0 bb}$ & $\hat{g}_{H^0 tt}$ & $\hat{g}_{H^0 bb}$ &  $\hat{g}_{A^0 tt}$ & $\hat{g}_{A^0 bb}$  & $g_{A^0 Zh^0}$ & $g_{A^0 ZH^0}$ & $\hat{g}_{ZZ H^0}$ & $\hat{g}_{ZZ h^0}$  \\  \hline
    B1 & 0.958 & 1.118 & -0.639 & 1.677 & 0.571  & 1.75 & -0.069 & -0.998 & -0.0689 & 0.998\\ 
    B2 & 1.108 & 1.108 & -0.684 & -0.684 & 0.833 & -0.833 & 0.141 & -0.990 & 0.141 & 0.990 \\ 
    B3  & 0.987 & 1.034 & -0.608 & 1.679 & 0.588 & 1.700 & -0.020 & -1.000 & -0.020 & 1.000  \\  \hline \hline  
   \end{tabular}
\end{center}
\caption{Normalised heavy--quark Yukawa couplings and Higgs $Z$ couplings for the
different 2HDM benchmarks defined in Table~3. Yukawa couplings are normalised to 
their SM counterparts as discussed in the text, while for the $A^0ZH^0$ and $A^0ZH^0$ couplings we show the proportionality constants of Eq.~3.3. }
\label{tab:couplingsYZ}
\end{table}

\subsection{2HDM results}
In this section we present our results for the three 2HDM benchmarks introduced in the previous paragraph. We start by considering the total cross section for each process, which is shown in Table \ref{tab:2HDM}. The heavy quark masses are again set to 173 and 4.75 GeV for top and bottom quarks, and the light Higgs mass to 125 GeV. The rest of the calculation details, such as the scale and PDF choices follow closely those of the SM calculation. We note here that where possible, we compared our results with the {\sc vh@nnlo } version described in \cite{Harlander:2013mla} and found very good agreement between the two implementations. 

\begin{table*}[t]
\renewcommand{\arraystretch}{1.3}
\begin{center}
    \begin{tabular}{ l | r | r | r  }
        \hline \hline
       & $gg\to Zh^0$
         & $gg\to ZH^0$
         & $gg \to ZA^0$\\
         \hline
          B1 &  
           113 $^{+30\%}_{-21\%}$ & 
686 $^{+30\%}_{-22\%}$ & 0.622 $^{+32\%}_{-23\%}$ \\ 
        B2 & 85.8 $^{+30.1\%}_{-21\%}$ & 1544 $^{+30\%}_{-22\%}$
& 0.869 $^{+34\%}_{-23\%}$ \\
B3 & 167 $^{+31\%}_{-19\%}$ & 0.891 $^{+33\%}_{-21\%}$ & 1325 $^{+28\%}_{-21\%}$ \\

\hline \hline
\end{tabular}
 \caption{\label{tab:2HDM} Cross sections (in fb) for gluon induced $Z$ Higgs associated production
   at the LHC at $\sqrt{s} = 14$~TeV for three 2HDM benchmarks. 
      The  uncertainties (in percent) refer to scale variations.  
      No cuts are applied to final state particles  and no Higgs or $Z$  branching ratios are included. }  
\end{center} 
\end{table*}

Before moving to the discussion of some differential results, we first comment on the results in Table \ref{tab:2HDM}. First we notice that  the cross-section for the $Zh^0$ process can be significantly enhanced. To be more precise, benchmark B3 leads to a cross section nearly twice the SM prediction, benchmark B1 to a 60\% enhancement, while B2 is gives a smaller $\sim$20\% increase. The main source of the increase in the cross-section is the presence of the resonant decay $A^0\to Zh^0$, which is kinematically allowed in all three scenarios. The relative change in the $Zh^0$ cross section is strongly correlated with the mass of the pseudoscalar and the value of the  $A^0Zh^0$ coupling. We remind ourselves that this coupling is proportional to cos$(\beta-\alpha)$,  {\it i.e.}, it tends to zero in the alignment limit. For all scenarios considered here,  its value remains small as seen in Table \ref{tab:couplingsYZ}. Consequently, it is not possible for this process to receive extremely large contributions from the resonance. This is in contrast with what we have seen in light Higgs pair production where the resonant decay of the heavy Higgs can lead to an enhancement of up to a factor of 60 for the $gg \to h^0h^0$ cross section \cite{Hespel:2014sla}. 

The most interesting feature of Table \ref{tab:2HDM}, is the potential size of the cross section for the $ZH^0$ process. We find that this can exceed 1~pb when the pseudoscalar $A^0$ is sufficiently heavy to allow the resonant decay into the heavy Higgs and a $Z$. This has been noticed and discussed recently in \cite{Dorsch:2014qja}, as a signature for a cosmologically motivated 2HDM scenario. It is remarkable that even if the production threshold lies significantly higher, this process can lead to larger cross sections compared to the $Zh^0$. This is possible as the relevant coupling, $ZH^0A^0$, as shown in Table \ref{tab:couplingsYZ},  is not suppressed by the ``SM-like" light Higgs constraints. Despite the fact that the prospects for discovery depend strongly on the resulting decay products of the heavy Higgs, it is worth noting that even in the scenarios where $H^0$ decays predominantly into $b\bar{b}$, the current experimental searches for $ZH$ set a cut on the invariant mass of the $b\bar{b}$ pair close to the light Higgs mass and would therefore miss this signal. Finally, we note that the $ZA^0$ production cross section remains very small in the scenarios where the $A^0$ is heavier than $H^0$, but can reach the picobarn level in a scenario such as benchmark B3, as a result of the inverted mass hierarchy.

\begin{figure}[h!]
\centering
\includegraphics[trim=2.5cm 0 0 0,scale=0.85]{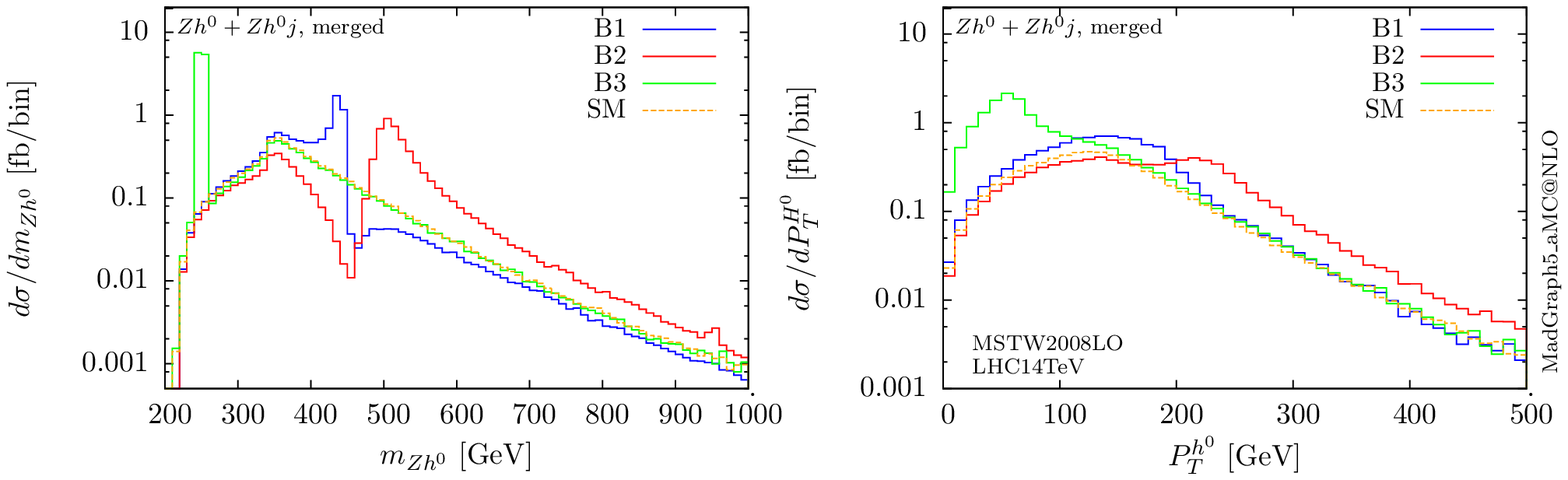}
\includegraphics[trim=2.75cm 0 0 0,scale=0.85]{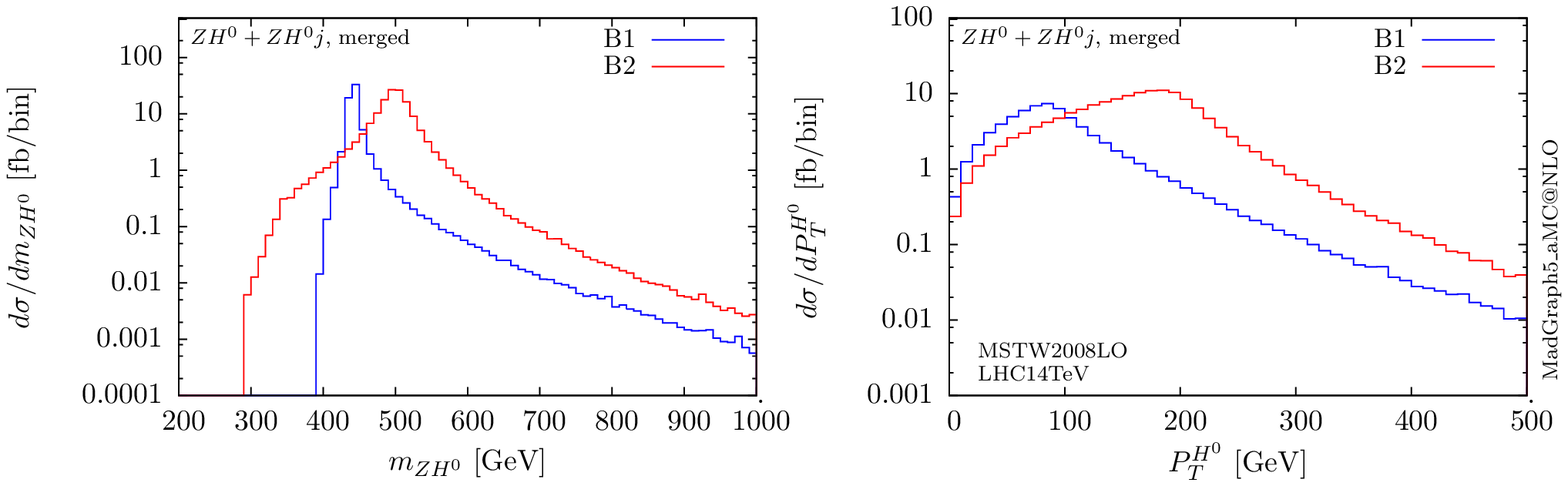}
\includegraphics[trim=2.9cm 0 0 0,scale=0.85]{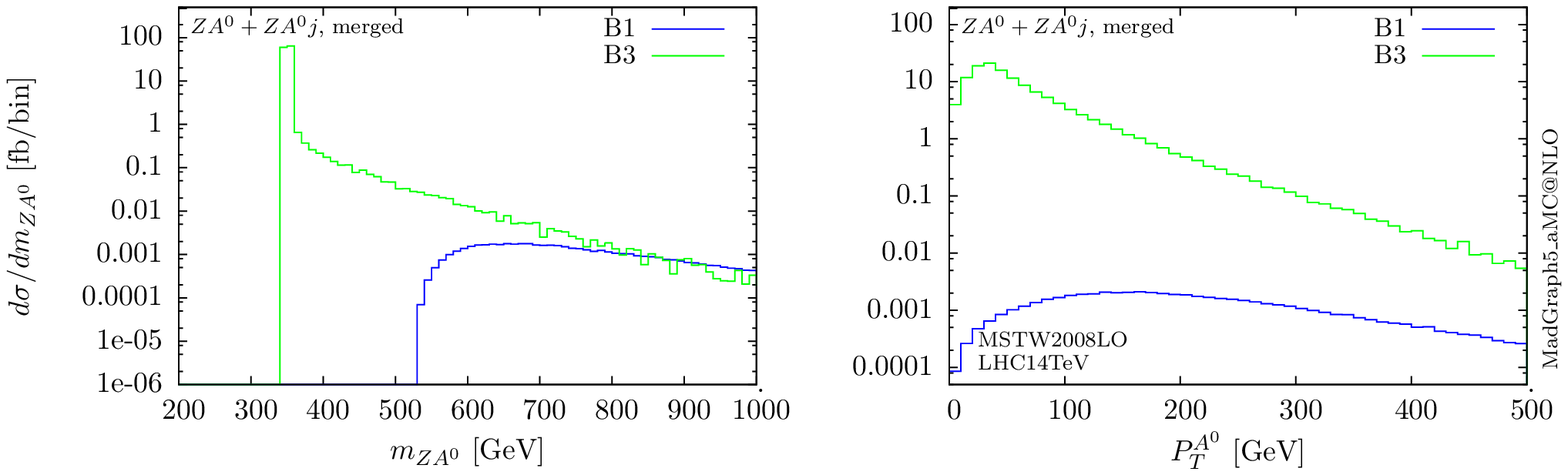}
\caption{Differential distributions for gluon induced $Z h^0$, $Z H^0$ and $Z A^0$ production at $\sqrt{s}=14$ TeV for the three 2HDM benchmarks and comparison with the SM. Left: Invariant mass of the $Z \phi$ system. Right: Transverse momentum of the Higgs.}
\label{zh1}
\end{figure}

Further interesting information on these processes can be extracted from the differential distributions. For brevity we present only those for the invariant mass of the system and the transverse momentum of the Higgs, but our setup is fully differential and any distribution can be plotted. We show these in Fig.~\ref{zh1}, for the cases in which the cross section is not negligible. The results shown here are obtained with merged samples of 0 and 1-jet matched to {\sc Pythia 8} for parton shower, in the same setup as that described in Section 2 for the SM. 

For the $Zh^0$ final state we also show the SM prediction for comparison. Resonance peaks arise in all scenarios for $Zh^0$, each time located at the mass of the pseudoscalar $A^0$. The sharpness of the peak varies with the mass of $A^0$, as heavier $A^0$ have larger widths going from 0.01 GeV for B3, to 7 GeV for B1 and 35 GeV in B2. We also notice various interesting interference patterns, clearly visible for benchmarks B1 and B2. The $A^0$-mediated diagram interferes with the SM-like amplitude, with the interference switching sign at $\sqrt{\hat{s}}=m_{A^0}$. Comparing scenarios B1 and B2, we see that the $Zh^0A^0$ couplings have opposite signs and therefore in one case the dip appears right before the resonance peak, and in the other right after. More subtle features are also visible in the plots away from the resonance peaks. These features can always be traced back to the 2HDM parameters and the value of the relevant couplings as shown in Table \ref{tab:couplingsYZ}. One such example is the fact that the B2 $m_{Zh^0}$ curve lies a bit lower than the SM one in the region below 350 GeV, which can be linked to the enhanced top Yukawa leading to a bigger box contribution. The box is in turn interfering destructively with the triangle leading to a smaller total amplitude for the $gg\to Zh^0$ process. 

For the $ZH^0$ process, only the two benchmarks that give measurable cross sections are shown. The plots shown for this process are dominated by the resonant decay of $A^0$. This is more obvious in the B1 curve as the resonant peak is closer to the threshold. Scenario B2 receives some non-negligible off-peak contributions from the $Z$ triangle and box diagrams, which in this case interfere constructively, as the $H^0$ top Yukawa sign is flipped. For B1 both the top Yukawa and $ZZH^0$ couplings signs are flipped, therefore the interference between triangle and box is destructive, and the result in the tails away from the resonant peaks, is suppressed compared to B2.

The situation is less complicated for $ZA^0$ for which in B3 a resonance very close to the $m_Z+m_{A^0}$ threshold dominates the plots, while the cross sections for B1 and B2 are extremely suppressed as no resonant decay is kinematically allowed. Moreover the production of  a rather heavy $ZA^0$ pair probes the gluon luminosity at large partonic $x$ values and is therefore suppressed.  

\section{Conclusions}
Investigating the nature of the Higgs boson discovered at the LHC is a challenging task. While the results of the measurements undertaken so far show that the 125 GeV scalar agrees well with the SM prediction, there is still room for deviations from the SM and possibly an extended Higgs sector to be discovered at the LHC. The exploration of various Higgs production processes is of vital importance to exclude or confirm a non-minimal Higgs sector. 

An important process yet to be precisely measured at the LHC is the associated production of a Higgs with a $Z$ boson. In addition to the Drell-Yan type contributions, this process  acquires a gluon fusion component at NNLO, which proves to be of particular importance in the boosted regime.  In this work, we have reviewed the main features of the $gg\to ZH$ process, both at the matrix-element and cross-section level. We have examined the behaviour and the relative  importance of the $2\to 2$ and $2 \to 3$ matrix elements for the gluon induced component. We have found that in the high $p_T$ regions the $2 \to 3$ matrix elements behave in a  different way from the $2 \to 2$ ones and therefore have to be taken into account to provide accurate predictions for the differential distributions.  To achieve this, we have combined the two in a consistent way, by merging different jet multiplicity samples and matching them to a parton shower.

Our results have been obtained within the {\sc MadGraph5\_aMC@NLO} framework with the help of {\sc Pythia 8} for the parton shower. The ME+PS approach provides a more accurate description of the process compared to the parton shower alone. In particular, it significantly reduces the uncertainty associated with the shower scale choice. For observables such as the transverse momentum of radiated jets in the hard region, the prediction of the parton shower alone can be misleading as here the results are extremely sensitive to the shower parameters. We find that in the merged predictions this sensitivity is almost completely eliminated, with the shower uncertainty remaining well within the intrinsic QCD uncertainty due to the renormalisation and factorisation scale variations.

The reduction of the uncertainties associated with the SM prediction and especially the accurate description of differential distributions is crucial for searches for beyond the SM scenarios. One scenario that the LHC aims to explore is the 2HDM. In this paper, we have also provided predictions for the gluon fusion component of the $Z\phi$ associated production in the 2HDM. Following the same setup as in the SM, we have presented our predictions for three representative 2HDM benchmarks. We have considered all three neutral Higgs bosons, presenting results for the cross sections and the differential distributions. 

In the production of the light Higgs in association with a $Z$, large enhancements can be achieved compared to the SM prediction if the resonant decay of the pseudoscalar $A^0$ is kinematically allowed. Moreover, interference patterns arise between the additional diagrams and the SM-like ones, leading to interesting features in the differential distributions. The resonant production of a $H^0Z$ pair also becomes important as the $H^0ZA^0$ coupling is not suppressed, leading to large cross sections for $gg \to ZH^0$ if the pseudoscalar $A^0$ is heavier then $H^0$. Finally in scenarios where the pseudoscalar $A^0$ is lighter than the heavy Higgs, $gg\to H^0\to ZA^0$ production is allowed and leads to large cross sections in still-to-be-excluded scenarios.

\section*{Acknowledgements}
We would like to thank David L\'opez-Val for his assistance with the 2HDM benchmarks, Stefan Liebler for a private version of {\sc vh@nnlo} with the 2HDM implementation and Ambresh Shivaji and Olivier Mattelaer for useful discussions.   This work has been performed in the framework of the ERC Grant No. 291377 ``LHCTheory''
and has been supported in part by the European Union as part of the FP7 Marie Curie Initial Training Network MCnetITN  (PITN-GA-2012-315877) and by the National Fund for Scientific Research (F.R.S.-FNRS Belgium) under a FRIA grant.

\bibliographystyle{JHEP}
\bibliography{zhbib}
\end{document}